\documentclass[a4paper,12pt]{article}
\usepackage[frenchb]{babel}
\usepackage[applemac]{inputenc}
\usepackage{lmodern,textcomp} 
\usepackage{setspace}
\usepackage{a4}
\usepackage{multicol}
\usepackage[pdftex]{graphics,graphicx}
\usepackage{amsfonts,amssymb,amsmath} 
\usepackage{wrapfig}
\usepackage{pifont} 
\usepackage{tikz} 
\usetikzlibrary{arrows,calc}
\usetikzlibrary{shapes,mindmap}
\newtheorem{definition}{D\'{e}finition}[section]

\newtheorem{rem}{Remarque}[section]
\begin{document}

\title{\Huge \textbf{La machine {\Huge{$\alpha$}} : modèle générique pour les algorithmes naturels}}
\author{Marc Bui\thanks{CHaRt-EA4004, Université paris 8 \& EPHE marc.bui@gmail.com}; Michel Lamure\thanks{EA 4128 Santé Individu Société, Université Lyon 1-UFR d'odontologie-11 rue Guillaune Paradin 69372 lamure@universite-lyon1.fr} \\  Ivan Lavall\'{e}e\thanks{CHaRt-EA4004, Université paris 8 \& EPHE ivan.lavallee@gmail.com}}
\date{15 avril 2013}
\maketitle
\vspace{-0.5cm}
\hrulefill

\medskip 
\noindent { \textbf{R\'{e}sum\'{e} :}}

{\small \bf Jusqu'ici, suite aux travaux de A.M.Turing \cite{Tur36}, les algorithmes ont été vus comme l'abstraction à partir de laquelle on pouvait écrire des programmes pour des ordinateurs dont le principe était lui-même issu du concept théorique de machine de Turing.

Nous partons ici du constat que les \textit{algorithmes naturels} ou plutôt les \textit{algorithmes de la nature}, massivement parallèles, autoadaptatifs et auto reproductibles,  dont on ne sait pas comment ils fonctionnent réellement, ni pourquoi, ne sont pas aisément spécifiés par le modèle théorique actuel de Machine de Turing Universelle, ou de Calculateur Universel; en particulier les aspects de communications, de règles évolutives, d'événements aléatoires, à l'image du code génétique, ne sont pris en compte que par ajout d'artifices à la théorie. Nous nous proposons ici de montrer comment aborder ces problèmes en repensant le modèle théorique. Nous proposerons un modèle d'algorithme, appelé ici \textit{machine-{\large$\alpha$}} qui contient et généralise les modèles existants. 
}\\
\rule[.5pt]{7cm}{1pt} \\
{\small \textbf{Mots clés:} Algorithme, algorithmes naturels, machine de Turing, calculateur universel, communications, généricité.}

\vspace*{3.5mm}
{\centering  \rule[.75pt]{5cm}{1pt} \par}

\vspace*{3.5mm}
\noindent {\textbf{Abstract :}}

{\small \bf So far, following the works of A.M. Turing, the algorithms were
considered as the mathematical abstraction from which we could write
programs for computers whose principle was based on the theoretical
concept of Turing machine.
We start here from the observation that natural algorithms or rather
algorithms of the nature which are massively parallel, autoadaptative
and reproductible, and for which  we do not know how they really work,
nor why, are not easily specified by the current theoretical model of Universal Turing machine, or Universal Computer. In particular the aspects of communications, evolutionary rules (rulers), random (unpredictable) events, just like the genetic code, are taken into account only by subtleties which oblige to break the theory. We shall propose one \textit{universal model} of algorithm called \textit{machine-{\large$\alpha$}} which contains and generalizes the existing models.
}\\
\rule[.5pt]{7cm}{1pt} \\
{\small \textbf{Key words:} Algorithm, natural algorithms, Turing machine, universal calculateur, communication, genericity.} \\
 %
 {\center \rule[.75pt]{5cm}{1pt} \par}
\section{Introduction}
\begin{figure}[h]
\begin{tikzpicture}
\draw[dashed] (0,0) circle (3.3) ;
\draw[fill=blue!50] (.75,1.5) -- (1.25,1.5) -- (1,2) -- cycle;%
\draw[fill=blue!50] (-.75,.5) -- (-1.25,.5) -- (-1,1) -- cycle;%
\draw[fill=blue!50] (-2.75,1.5) -- (-3.25,1.5) -- (-3,2) -- cycle;
\draw[fill=blue!50] (2.5,-2.5) -- (3,-2.5) -- (2.75,-2) -- cycle;
\draw[fill=blue!50] (-1.75,-1.2) -- (-2.25,-1.25) -- (-2,-.75) -- cycle;
\draw[fill=green!50] (-.25,-.5) -- (.25,-.50) -- (0,0) -- cycle;
\draw [color=green,line width=1.5pt,->] (0,-.1) -- (1,1.5); %
\draw [color=green,line width=1.5pt,->] (-.2,-.3) -- (-1,.5); %
\draw [color=green,line width=1.5pt,->] (-.2,-.3) -- (-2,-1); %
\draw [color=red,line width=3pt,->] (.1,-.25) -- (2,-1.5); %
\draw (-.5,-3.8) node[above] {Maintenir une distance}; 
\draw (-.5,-4.2)node[above]{minimale avec les autres};
%
%
\draw[dashed] (9,0) circle (3.3) ;
\draw[fill=blue!50] (9.75,1.5) -- (10.25,1.5) -- (10,2) -- cycle;%
\draw[fill=blue!50] (8.25,.5) -- (7.75,.5) -- (8,1) -- cycle;%
\draw[fill=blue!50] (6.25,1.5) -- (5.75,1.5) -- (6,2) -- cycle; 
\draw[fill=blue!50] (9,-2.5) -- (9.5,-2.5) -- (9.25,-2) -- cycle;
\draw[fill=blue!50] (10.5,0) -- (11,0) -- (10.75,.5) -- cycle;
\draw[fill=blue!50] (7.25,-1.2) -- (6.75,-1.25) -- (7,-.75) -- cycle;
\draw[fill=black!100](9,.2) circle(.2);
\draw[fill=green!50] (9.25,-.5) -- (8.75,-.50) -- (9,0) -- cycle;
\draw [color=red,line width=3pt,->] (8.75,-.25) -- (8,-.25); %
\draw (9.5,-3.8) node[above] {Se déplacer vers le centre perçu.}; 
\draw (9.5,-4.3)node[above]{\textbf{SI} obstacle de face \textbf{ALORS}}; \draw (9.5,-4.9)node[above]{prendre la tangente};
\end{tikzpicture}\\
%
\vspace*{-1cm}
\caption{Algorithme \textit{naturel} des Boïds de \cite{Rey87}\label{boids}}
\end{figure}
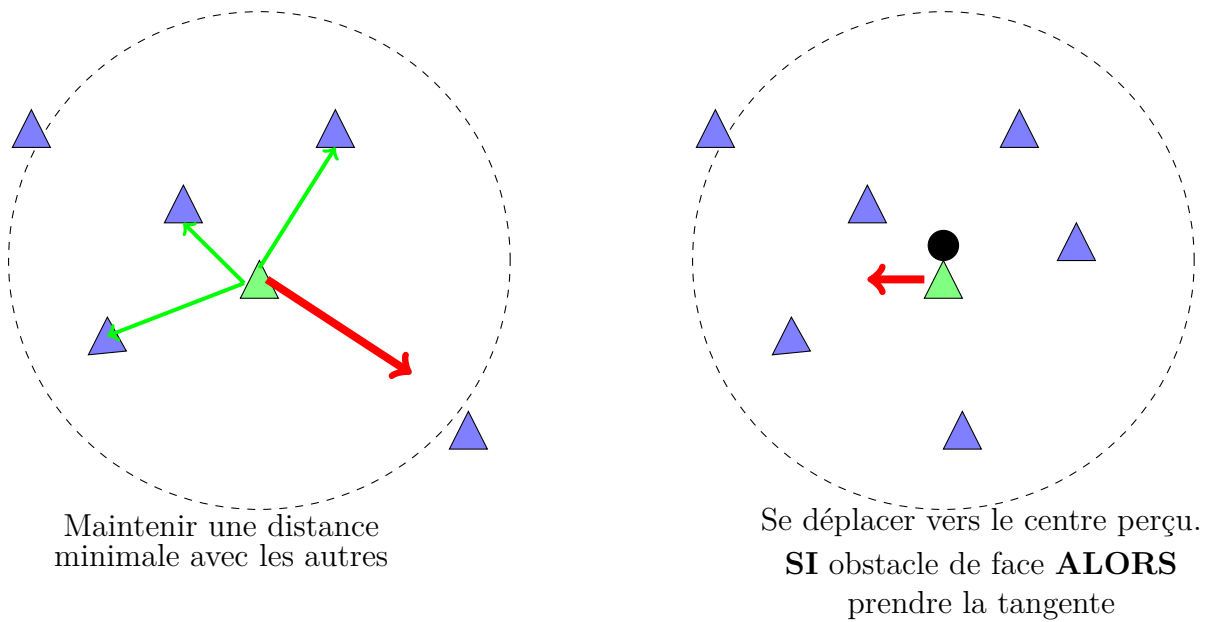

Qu'y-a-t-il de commun entre la sélection naturelle décrite par Darwin \cite{Dar72}, l'arbre phylogénétique des mamifères \cite{CTM06},  un vol d'oiseaux auto-organisé \linebreak
\cite{Rey87}, la synchronisation de Kuramoto \cite{Kur75}, ou les fourmis de Langton, \cite{Lan86}. La disparition massive des grands dinosaures suite à la modification de leur environnement dû à un événement contingent extérieur  (chute d'une comète, d'une météorite…) ressortit  de même aux algorithmes \textit{naturels} (voir \linebreak\cite{Cha12-a}), mais dans ce dernier cas, c'est {\em \textbf{l'influence}} du contexte qui est déter\-minante. On trouvera une illustration de ce qu'on appelle {\em Algorithme naturel} en \S{}\ref{algonat} figure \ref{boids}.

Quelle algorithmique pour en rendre compte ? Si on peut voir effectivement ces  phé\-nom\-ènes naturels à travers le prisme des algorithmes en tant que tels, certains phénomènes comme l'influence du contexte, l'imprévisibilité, le temps ou la communication nous conduisent à revenir sur le concept fondateur lui-même.

 Si Darwin a bien  formulé et étayé l'hypothèse selon laquelle toutes les espèces vivantes ont évolué au cours du temps à partir d'un seul ou quelques ancêtres communs grâce au processus connu sous le nom de sélection naturelle\footnote{Nous reviendrons sur la signification de ce concept.}, processus non formalisé, la génétique a montré que cela se traduisait par des modifications dans l'ADN, l'ARN et les protéines. La modélisation en peut être faite à partir de la théorie algorithmique revisitée en faisant appel aux travaux d'Andreï Kolmogorov (voir \cite{LiVi97,SaBo11}). 

Les ordinateurs et leurs algorithmes ont révolutionné la façon d'aborder la simulation et le calcul, permettant d'en avoir une vision incrémentale et expérimentale. Il s'agit là d'un outil puissant permettant non seulement d'aborder des domaines nouveaux, mais surtout de les aborder différemment qu'avec les outils mathématiques classiques, on élargit là le champ d'investigations possibles. En particulier l'aspect évolutif de la vie ; la théorie de Darwin ouvre à l'algorithmique un champ nouveau de même nature que l'a été ce qu'on a appelé \og{}\textit{intelligence artificielle} \fg{}. On examinera en particulier les questions suivantes:
\begin{enumerate}
\item comment simuler algorithmiquement la morphogenèse, de la bactérie à l'Homo sapiens, au niveau principiel, sans avoir à entrer dans l'étude des processus biologiques, comme \cite{Kau69} qui utilise pour ce faire des réseaux booléens aléatoires;
\item comment rendre compte de la constitution de l'arbre phylogénétique et en particulier, les variations dans les espèces;
\item comment s'opèrent certaines coordinations ou coopérations entre individus, comme la constitution d'un vol d'oiseaux migrateurs partant en migration (des boids dans le vocabulaire de \cite{Rey87}) ou la synchronisation des lucioles dans le sud est asiatique telle que décrite par Sir \textit{Francis Drake} en 1577 et rapportée par \cite{Pre00};
\item comment faire intervenir le contexte comme par exemple une chute de météorites ou une inversion des pôles magnétiques ou autres phénomènes affectant l'environnement ?
\end{enumerate}
\subsection{Un algorithme naturel\label{algonat}}
%
Le schéma algorithmique ci-dessous, illustre ce que nous appelons ici {\em algorithme naturel} il simule la formation d'un vol d'oiseaux appelés boïds par l'auteur. 

\bigskip
%
%
\subsubsection{La spécification}
Il suffit de quatre règles pour spécifier ce comportement algorithmique, comme spécifié sur les dessins:
\begin{enumerate}
\item maintenir une \textit{distance minimale} avec les autres;
\item se déplacer vers le centre perçu (le barycentre);
\item \textbf{SI} obstacle de face \textbf{ALORS} prendre la tangente;
\item \textbf{SI} obstacle percuté \textbf{ALORS} repartir plus vite
\textbf{TANT-QUE} loin du groupe .
\end{enumerate}

La simulation informatique montre que les boïds finissent par s'organiser de même façon qu'un vol réel d'oiseaux migrateurs.
\section{Nos objectifs}
Il s'agit de développer un cadre de travail pour :
\begin{itemize}
\item[\textbullet] la modélisation des algorithmes tels que déjà connus;
\item[\textbullet] spécifier les algorithmes des phénomènes émergents. En effet, ces derniers ont une dynamique qui n'est pas représentable par des variables d'état seules et nécessite un abord algorithmique. Plus précisément, il nous faut rendre compte des {\em algorithmes de la nature}, particulièrement :
	\begin{itemize}
	\item des systèmes d'influence,
	\item du concept de proximité,
	\item de la synchronisation de comportements,
	\item des modifications du génotype et du phénotype et conséquemment du concept de {\em sélection naturelle};
	\end{itemize}
\item[\textbullet] formaliser le concept de {\em Machine de Turing non déterministe\footnote{\`A ne pas confondre avec le modèle RAM, Random Access Memory.}} ou MTND et rester cohérent avec les concepts non déterministes propres à la théorie de la complexité calculatoire.\cite{LiVi97, SaBo11};
\item[\textbullet] inclure les résultats déjà connus tels la machine de Turing et le calculateur universel et rester cohérent avec les résultats afférents.
\end{itemize}
\section{Le contexte d'étude}
\begin{itemize}
\item les algorithmes en tant que tels;
\item les réseaux, c'est-à-dire des algorithmes qui communiquent entre eux;
\item l'environnement dans lequel tout se passe.
\end{itemize}
\subsection{Qu'est-ce qui est nouveau ?}
En 1945 Turing produit un rapport intitulé \og {\em Proposed Electronic Calculator}\fg{}; c'est son projet d'ordinateur ACE (Automatic Computing Engine) qu'il présente ainsi : \footnote{\em “The idea behind digital computers may be explained by saying that these machines are intended to carry out any operations which could be done by a human computer. The human computer is supposed to be following fixed rules; he has no authority to deviate from them in any detail. We may suppose that these rules are supplied in a book, which is altered whenever he is put on to
a new job. He has also an unlimited supply of paper on which he does his calculations.”}

{\em L'idée qui préside à la fabrication des calculateurs numériques est que ces machines sont destinées à effectuer toutes les opérations qui pourraient être faites par un calculateur humain. On suppose que le calculateur humain suit des règles fixes; il n'a aucune autorité pour en dévier dans aucun détail. Nous pouvons supposer que ces règles sont fournies dans un livre, qui change pour tout nouveau calcul. Le calculateur a une provision illimitée de papier sur lequel il fait ses calculs\footnote{traduction libre I.L.}.}

Cette vision est celle du calculateur, en anglais "\textit{computer}\footnote{d'où le nom \og computer science\fg{}, lequel, traduit en français, signifie {\em science du calculateur} ce à quoi ne se réduit pas l'informatique...}" pas celle de l'ordinateur, elle ne contient pas les communications entre machines, ni la possibilité de taches, mites ou ratures et ajouts sur les pages du livre évoqué par A.M. Turing; c'est de ces dernières situations dont nous entendons traiter ici.

La situation décrite par Turing est un cas particulier d'une situation algorithmique telle qu'il n'y ait qu'un seul humain qui fasse le calcul qu'il n'y ait pas de tache sur le livre, qu'on n'en ait pas arraché des pages  ou raturé des paragraphes ni que personne ne passe à côté avec une pipe ou une tasse de café, ou un stylo qui fuit, qui laisse tomber des brandons ou des gouttes qui font des marques sur le livre, ni personne qui annote le livre. 

Nous entendons ici traiter ces problèmes en lesquels une tache de café peut changer l'histoire; ce qui change avec cette algorithmique nouvelle, c'est la prise en compte de la communication, du parallélisme, ainsi que de l'aspect probabiliste :
\subsubsection{Les communications}
Les objets simulés par les algorithmes de la nature ou les bio-algorithmes communiquent entre eux, que ce soit chimiquement (phéromones), par la voix (aboiements, hululements,\ldots), par la vue, etc . De même, dans les réseaux d'ordinateurs, ceux-ci communiquent et ces communications ont des conséquences sur l'évolution ultérieure du calcul.

On retiendra ici quatre grands types de communications:
\begin{enumerate}
\item par réception de message ou \og communication passive\fg{};
\item par envoi de messages ou par écriture dans des registres (dans le cas d'ordinateurs);
\item par synchronisation;
\item par le contexte\footnote{On entend par là l'environnement dans lequel \og baigne \fg{} le système, champ électromagnétique, eau air, milieu igné...}. Par exemple, sans envoyer explicitement un message, une entité peut agir sur son environnement, et ainsi obliger une autre entité du même système à modifier le sien, c'est une communication passive. Les systèmes ainsi régis sont appelés ici {\em systèmes à influences}.
\end{enumerate}
\medskip
La simulation de ces communications peut se faire en considérant qu'il s'agit de messages, mais en distinguant deux types de protocoles de communication. 
\begin{itemize}
\item l'un où sont re\c cus et traités des messages non sollicités (ex: accident,  chute de météorites \ldots);
\item un qui traite des réponses à des messages; par exemple, pour connaître le voisinage, il faut regarder, ceci est modélisable par la réception d'un signal de capteur ou par un message interrogatif, la réponse en étant ce qui est vu.
\end{itemize}
\subsubsection{Le caractère distribué massif}
Des millions, voire des milliards, d'organismes vivants et d'objets sont concernés à un instant donné, mais pas tous de la même façon. 

C'est dans ce contexte que se posent les problèmes dits de \textit{concurrence}, de  \textit{coopération}, de  \textit{communication} et de  \textit{synchronisation}. Ni la Machine de Turing Universelle (MTU), ni le Calculateur Universel (CU) ne permettent de rendre compte ni de poser correctement ces problèmes.

Dans ces modèles théoriques, le temps par exemple est endogène au modèle, les possibilités de communications ne sont pas prises en compte et l'aspect aléatoire est exogène au modèle. L'algorithmique distribuée permet de disposer de nombreux outils théoriques, en particulier dans la prise en compte des communications, mais il s'agit là de modèles spécifiquement dédiés aux réseaux d'ordinateurs et qui n'ont pas la souplesse nécessaire à la prise en compte des phénomênes naturels.
\subsubsection{Influence du contexte}
Les algorithmes de la nature évoluent, par définition, dans un contexte changeant, que ce soit électromagnétique, chimique ou social, ne serait-ce aussi que par la communication. Ce caractère changeant n'est pas toujours prévu. Il ressortit à une modélisation probabiliste.

Nous montrons comment rendre compte de manière endogène au modèle de cette influence, ce qui permet de donner aussi un aspect plus formel au concept de {\em sélection naturelle}.
\paragraph{Le caractère probabiliste du contexte}
Le contexte est relativement imprévisible et soumis à catastrophes (impacts de corps célestes, pollution chimique, volcanisme, changements électromagnétiques‚ \ldots) et la façon idoine permettant de modéliser ces changements de contextes ou de modifications, c'est d'utiliser la théorie des probabilités. Dans les modèles classiques de spécification des algorithmes (MTU ou CU) cet aspect est exogène au modèle. Notre modèle le rend endogène. Il en est de même pour l'artifice introduit par certains auteurs en théorie algorithmique, celui de {\em machine de Turing à oracle} ou tout simplement d'\textit{oracle}, objet totalement exogène au modèle que nous rendons aussi ici endogène. 

.

\section{Rappels, définitions et vocabulaire}
Cette section nous permet de préciser ce qu'est une Machine de Turing \textit{particulière} (désormais MT), la Machine de Turing Universelle (désormais MTU), le Calculateur Universel (désormais CU), ainsi que le vocabulaire afférent utilisé:
%

\subsection{Modèle classique d'algorithme\label{algo}}
Le modèle théorique classique d'algorithme est la machine de Turing \cite{Tur36} qui a permis de fonder formellement les concepts de calcul, solution, algorithme, schème et programme.
\subsubsection{La machine de Turing \label{MT}}
 
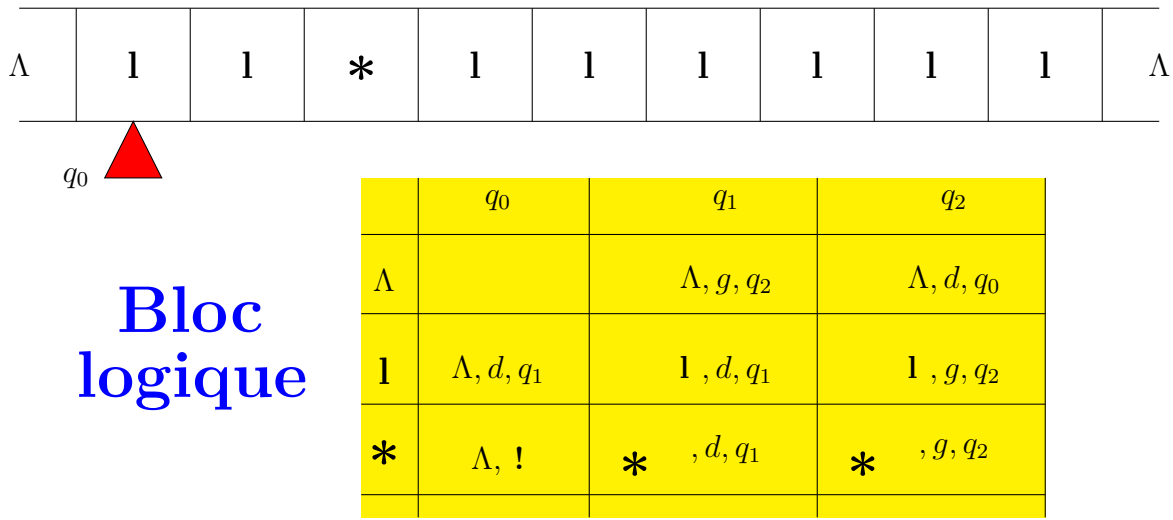
\begin{figure}[h]
\hrulefill
\begin{center}%
\medskip
\begin{tikzpicture}[scale=1.5]
\draw(0.5,6)--(0.5,5);
\draw(1.5,6)--(1.5,5);
\draw(2.5,6)--(2.5,5);
\draw(3.5,6)--(3.5,5);
\draw(4.5,6)--(4.5,5);
\draw(5.5,6)--(5.5,5);
\draw(6.5,6)--(6.5,5);
\draw(7.5,6)--(7.5,5);
\draw(8.5,6)--(8.5,5);
\draw(9.5,6)--(9.5,5);

%
%
%
%
\fill[yellow] (3,1.5) rectangle +(6,3) (7,0); 
%

\draw(3, 4)--(9,4); 
\draw(3.5,4.5)--(3.5,1.5);


%
\node[text centered] at(3.2,3.6){\bf $\Lambda$};
\node[text centered] at(3.2,2.8){\bf \large l};
\node[text centered] at(3.2,2){\bf \LARGE *};
\node[text centered] at(4.2,4.3){\bf $q_{0}$};
\node[text centered] at(6.2,4.3){\bf $q_{1}$};
\node[text centered] at(8.2,4.3){\bf $q_{2}$};
\node[text centered] at(6.2,3.6){\bf $\Lambda,g,q_{2}$};
\node[text centered] at(8.2,3.6){\bf $\Lambda,d,q_{0}$};
\node[text centered] at(4.2,2.8){\bf $\Lambda,d,q_{1}$};
\node[text centered] at(6.2,2.8){\bf l $,d,q_{1}$};
\node[text centered] at(8.2,2.8){\bf l $,g,q_{2}$};
\node[text centered] at(4.2,2.){\bf $\Lambda,$ !};
\node[text centered] at(5.4,1.9){\bf {\LARGE *}};
\node[text centered] at(6.2,2.1){\bf $,d,q_{1}$};
\node[text centered] at(8.2,2.1){\bf $,g,q_{2}$};
\node[text centered] at(7.4,1.9){\bf {\LARGE *}};
\draw(5,4.5)--(5,1.5);
\draw(3, 3.3)--(9,3.3); 
\draw(3, 2.5)--(9,2.5); 
\draw(3, 1.7)--(9,1.7); 
\draw(7,4.5)--(7,1.5);
\draw(9,4.5)--(9,1.5);
%
\draw(0,5)--(10,5);
\draw(0,6)--(10,6);
%
%
\draw[fill=red!100] (.75,4.5) -- (1.25,4.5) -- (1,5) -- cycle;
\node[text centered] at(.5,4.5){\bf $q_{0}$};
%
%
\node[text width=3cm,text centered] at(1.5,3){ \bf \Huge\color{blue}Bloc \\ \medskip logique};
\node[text centered] at(0,5.5){\bf $\Lambda$};
\node[text centered] at(1,5.5){\bf \large l};
\node[text centered] at(2,5.5){\bf \large l};
\node[text centered] at(3,5.4){\bf \LARGE *};
\node[text centered] at(4,5.5){\bf \large l};
\node[text centered] at(5,5.5){\bf \large l};
\node[text centered] at(6,5.5){\bf \large l};
\node[text centered] at(7,5.5){\bf \large l};
\node[text centered] at(8,5.5){\bf \large l};
\node[text centered] at(9,5.5){\bf \large l};
\node[text centered] at(10,5.5){\bf $\Lambda$};
\end{tikzpicture}
\vspace*{-2cm}
\caption{Machine de Turing de l'addition\label{T}}
\end{center}
\hrulefill
\end{figure}
%
La machine de Turing de la figure \ref{T} effectue l'addition de deux nombres représentés par des bâtonnets et séparés par un astérisque.

La position de départ est donnée ici par la position de l'index de {\em lecture/écriture} figuré par le triangle et l'état de la machine est $q_{0}$. On lit dans la table à double entrée (\textbf{l},$q_{0}$ $\longrightarrow \Lambda, d, q_{1}$) qui signifie, :{\em ayant lu le symbole {\em \textbf{l} }sur le ruban, alors que la machine est dans l'état $q_{0}$, écrire $\Lambda$ {\em (c'est-à-dire effacer ce qu'il y a dans la case)} décaler la tête de {\em lecture/écriture} d'une case vers la droite et mettre la machine dans l'état $q_{1}$}. Et ainsi de suite. 

\begin{definition} Une Machine de Turing   
(désormais MT) à un ruban est un quintuplet;
 \[ M=<\mathcal{Q}, \Sigma, q_0, t, F> \] tel que : 
\begin{itemize}

\item $\mathcal{Q} = \{ q_0, \ldots, q_n\}, n \in \mathbb{N}^*$ est un ensemble  fini d'états (ou alphabet intérieur) ; $q_0$ étant l'{\bf état initial} ; $F \subseteq {\mathcal{Q}}$ , $F$ étant l'ensemble des {\bf
états finaux}, c'est à dire pour un problème de décision:
\[ F = \lbrace q_{oui} ; q_{non} \rbrace ;\]
\item $\Sigma$ est l'{\bf alphabet} (alphabet extérieur), c'est un ensemble fini de symboles tel que ; $\mathcal{Q}$ et $\Sigma$ sont totalement disjoints, c'est-à-dire : $ \mathcal{Q} \cap \Sigma = \emptyset $ . De plus, $\Sigma$ contient toujours le symbole vide, noté $\Lambda$. Certains auteurs ajoutent un symbole particulier comme premier symbole signifiant comme par exemple $\rhd ;$ indiquant que là doit commencer la lecture des symboles sur le ruban mais si on donne la position de la tête de lecture-écriture en position de départ, ce symbole n'est plus nécessaire\footnote{Dans la littérature, on peut aussi trouver d'autres façons de formaliser le concept de machine de Turing, par exemple en caractérisant le symbole blanc ce qui conduit à distinguer l'alphabet de travail d'un  alphabet plus général. On décrit alors formellement une machine de Turing comme un septuplet:
\[M=<\mathcal{Q}, \Sigma, \Gamma, B, q_0, t, F>\]
 où $B$ est le symbole "blanc", tel que $B \in \Gamma$ et $B \notin \Sigma$; $\Gamma$ l'alphabet de travail (on a alors $\Sigma \subsetneq \Gamma$)};

\item $t$ est la {\bf fonction de transition} ;
\[t :\mathcal{Q} \times \Sigma \rightarrow \mathcal{Q} \times
\Sigma  \times \{G,D,N\}.\]
suivant les auteurs, on parle de \textit{schème} de la machine de Turing; $t$ peut être considéré en fait comme le "programme" de la machine. On utilise aussi le terme de {\em dérivation} pour une transition simple ou un {\em pas de calcul};
\item  $\{G,D,N\}$ est un alphabet de déplacement concernant la tête de lecture (ou le ruban) pour lequel:
	\begin{itemize}
	\item[$G$] signifie {\em décalage d'une case vers la gauche},
	\item[$D$] signifie {\em décalage d'une case vers la droite},
	\item[$N$] signifie {\em Neutre, c'est-à-dire, pas de décalage};
	\end{itemize}
\end{itemize}
\end{definition}
 
Cette présentation d'une MT est celle de ce qu'il est convenu d'appeler une MT particulière. En effet, il faut une MT pour l'addition, une pour la multiplication \ldots Pour chaque opération particulière, il faut définir une MT particulière, c'est-à-dire une fonction de transition, un alphabet intérieur (ou alphabet d'état) et un alphabet extérieur particularisés pour résoudre le problème considéré. Se pose alors la question de disposer d'une MT capable d'émuler tout MT particulière, c'est-à-dire une MT universelle.
\subsection{La machine de Turing universelle\label{MTU}}
 La théorie algorithmique est basée sur le concept théorique de {\em Machine de Turing Universelle} (désormais MTU) ; nous utiliserons aussi ici une façon particulière de représenter une MTU, à savoir le {\em calculateur universel} (\ref{CU}). 
 
 Comme écrit précédemment, une MTU doit être capable de simuler toute MT particulière. Comme les MT particulières, la MTU est constituée d'un ruban, d'une tête de lecture/écriture, d'un bloc logique (voir figure \ref{MTU}).


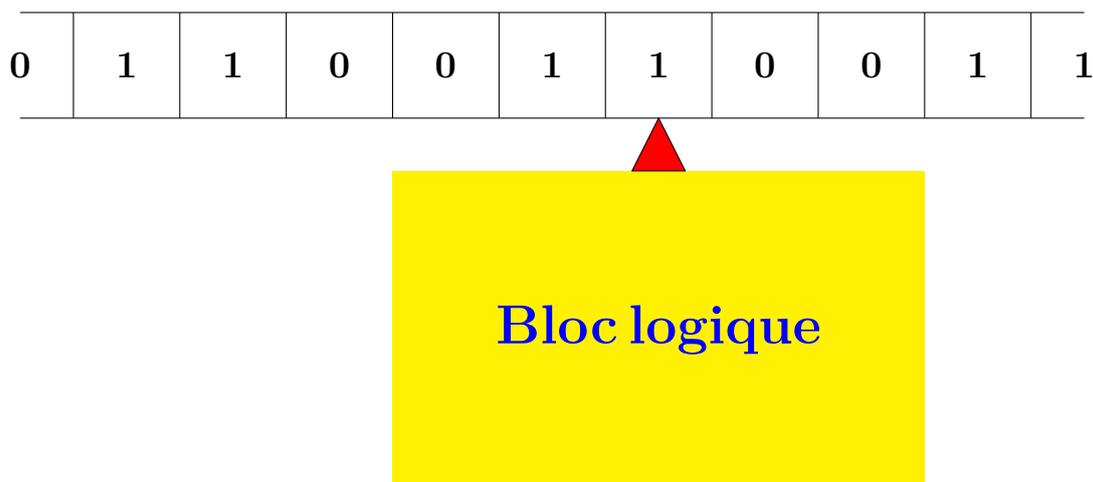
\begin{figure}[h]
\hrulefill
\begin{center}
\medskip
\begin{tikzpicture}[scale=1.4]
\draw(0.5,6)--(0.5,5);
\draw(1.5,6)--(1.5,5);
\draw(2.5,6)--(2.5,5);
\draw(3.5,6)--(3.5,5);
\draw(4.5,6)--(4.5,5);
\draw(5.5,6)--(5.5,5);
\draw(6.5,6)--(6.5,5);
\draw(7.5,6)--(7.5,5);
\draw(8.5,6)--(8.5,5);
\draw(9.5,6)--(9.5,5);
%
%
%
%
\fill[yellow] (3.5,1.5) rectangle +(5,3) (7,0); 
%
%
\draw(0,5)--(10,5);
\draw(0,6)--(10,6);
\draw[fill=red!100] (5.75,4.5) -- (6.25,4.5) -- (6,5) -- cycle;
%
%
\node[text width=5cm,text centered] at(6,3){ \bf \LARGE \color{blue}Bloc logique};
\node[text centered] at(0,5.5){\bf \large 0};
\node[text centered] at(1,5.5){\bf \large 1};
\node[text centered] at(2,5.5){\bf \large 1};
\node[text centered] at(3,5.5){\bf \large 0};
\node[text centered] at(4,5.5){\bf \large 0};
\node[text centered] at(5,5.5){\bf \large 1};
\node[text centered] at(6,5.5){\bf \large 1};
\node[text centered] at(7,5.5){\bf \large 0};
\node[text centered] at(8,5.5){\bf \large 0};
\node[text centered] at(9,5.5){\bf \large 1};
\node[text centered] at(10,5.5){\bf \large 1};
\end{tikzpicture}
\vspace*{-2cm}
\caption{Machine de Turing Universelle\label{MTU}}
\end{center}
\hrulefill
\end{figure}
Du point de vue conceptuel, il n'y a alors pas de différence entre données et schème de MT. La séquence de symboles sur le ruban concerne tant les données que le code (schème) de la MT particulière simulée. Pour ce faire, il faut que le bloc logique de la MTU puisse distinguer les symboles qui ressortissent aux données du calcul à effectuer de ceux des symboles qui codent l'algorithme à exécuter ainsi que ceux des symboles du résultat.
 
Une formulation pratique de la MTU est de distinguer les différentes séquences de
symboles du ruban. L'alphabet extérieur de la MTU sert à coder :
\begin{enumerate} 

\item \textbf{le schème} de la machine (l'algorithme) 
particulière considérée ; 

\item \textbf{l'instance particulière} de données sur laquelle il faut
exécuter le programme ; 

\item \textbf{le résultat} du calcul de la machine elle même.

\end{enumerate} 
\subsubsection{Codage} 
Le caractère \textit{universel} de la machine est obtenu par le fait que la communication avec l'extérieur (l'environnement) se fait par l'intermédiaire du ruban, et il faut que tout passe par ce ruban, les données, le schème qu'on appelle alors le programme, et l'écriture du résultat. Il faut alors distinguer ce qui ressortit dans le codage au schème, à l'instance de données, et au résultat. Le codage binaire permet de faire cela, en distinguant les différents langages par codage. Une possibilité en est :
\begin{itemize}
\item chaque caractère du langage extérieur peut être codé $1$ suivi d'un nombre pair de zéros supérieur à $3$ et se terminant par $1$, ainsi $100001$ est un symbole du langage considéré;
\item  chaque caractère décrivant l'espace d'états (ou langage intérieur)  peut être codé $1$ suivi d'un nombre impair de zéros supérieur à 3 et se terminant par $1$, ainsi $1000001$ est un symbole du langage considéré;
\item les chaînes $101\; ; \; 1001\; ; \; 10001$ codant l'alphabet de mouvement, par exemple {\em décalage à droite, décalage à gauche, stagnation}.
\end{itemize}
La séparation entre les différents caractères tient à ce type de codage. En effet, une \textit{séquence} de la MTU se présente alors sous la forme suivante :

\[\ldots 1000000110011000001\ldots\] 

les différents caractères des différents alphabets sont alors séparés et reconnus par la succession de deux $1$, le début et la fin de la séquence n'étant eux marqués que par un seul caractère $1$.
\subsection{Machine de Turing à plusieurs rubans}
On peut sans perte de généralité considérer un ruban pour 
chacune de ces séquences de symboles, ce qui nous conduit à 
utiliser le concept de machine à plusieurs rubans.

Cette fa\c con de présenter les choses a un avantage, c'est celui de précisément distinguer, pour des raisons de lisibilité théorique, le programme proprement dit du résultat et des calculs intermédiaires, permettant ainsi de préciser les attributions et rôle de chacun.

Nous considérerons ici qu'il y a trois rubans.

Rien n'est dit ici bien sûr sur le \og {\em programme système} \fg{} de la machine universelle elle-même. Il suffit pour dire qu'il existe que les codages se font avec des alphabets finis .
\subsection{Calculateur universel\label{CU}}
Un \textit{calculateur universel} \cite{Lav10,SaBo11}, (désormais CU) est une variante de MTU à trois rubans dont on a spécialisé les rubans, c'est-à-dire:
\begin{itemize} 
\item[\ding{43}] sur le premier ruban se trouve le \textit{programme}. Ce ruban est infini à droite (respectivement à gauche) et rempli de $0$ et de $1$ codant le programme, la machine peut en lire le contenu et ne peut faire que ça sur ce ruban. Au départ ce ruban possède une écriture, le programme à exécuter. La tête de lecture ne se déplace que dans une seule direction, d'une case (et d'une seule) à chaque \textit{item} de temps sans possibilité de retour en arrière et sans non plus de possibilité de stagnation. Ce sont ces deux dernières propriétés qui font la caractéristique du calculateur universel par rapport à la MTU;
\item[\ding{43}] le deuxième ruban est un ruban dit {\em de travail}, pour les calculs intermédiaires sur lequel la machine peut lire,
effacer, écrire et décaler la tête de lecture/écriture d'une case (et d'une seule) à chaque item de temps vers la
droite, la gauche, ou rester stable, ce ruban est infini à
droite et à gauche. Au début du calcul, ce ruban comprend une
suite {\bf finie} de 0 et de 1, et des vides partout
ailleurs ; ce ruban contient, codée en $0$ et $1$ la donnée (l'instance)  $s$ du calcul; 
\item[\ding{43}] le troisième ruban est le ruban de \textit{résultat}
entièrement vide au début du calcul, une tête d'écriture
vient écrire le résultat (codé uniquement en $0$ et $1$), le ruban est infini à droite, la tête d'écriture ne peut se déplacer que vers la droite, uniquement et elle ne peut qu'écrire. 
\end{itemize} %
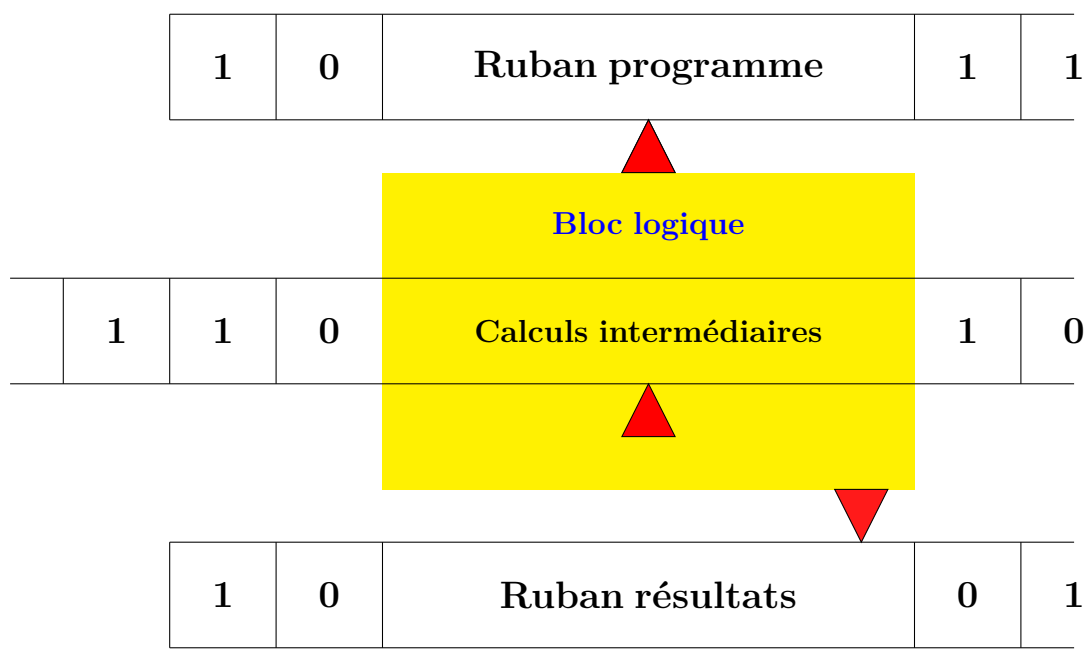
\begin{figure}[h]
\hrulefill
\begin{center}%
\medskip
\begin{tikzpicture}[scale=1.4]
\draw(1.5,0)--(10,0);
\draw(1.5,1)--(10,1);
\draw(1.5,1)--(1.5,0);
\draw(2.5,1)--(2.5,0);
\draw(3.5,1)--(3.5,0);
\draw(8.5,1)--(8.5,0);
\draw(9.5,1)--(9.5,0);
%
%
\draw(1.5,5)--(10,5);
\draw(1.5,6)--(10,6);
\draw(1.5,6)--(1.5,5);
\draw(2.5,6)--(2.5,5);
\draw(3.5,6)--(3.5,5);
\draw(8.5,6)--(8.5,5);
\draw(9.5,6)--(9.5,5);
%
%
\draw[fill=red!90] (5.75,4.5) -- (6.25,4.5) -- (6,5) -- cycle;
\fill[yellow] (3.5,1.5) rectangle +(5,3) (7,0); 
%
%
\draw(0,2.5)--(10,2.5);
\draw(0,3.5)--(10,3.5);
\draw(0.5,2.5)--(0.5,3.5);
\draw(1.5,2.5)--(1.5,3.5);
\draw(2.5,2.5)--(2.5,3.5);
\draw(9.5,2.5)--(9.5,3.5);
\draw[fill=red!90] (7.75,1.5) -- (8.25,1.5) -- (8,1) -- cycle;
%
\draw[fill=red!100] (5.75,4.5) -- (6.25,4.5) -- (6,5) -- cycle;
\draw[fill=red!100] (5.75,2) -- (6.25,2) -- (6,2.5) -- cycle;%
%
%
\node[text width=3.5cm,text centered] at(6,4){ \bf \color{blue}Bloc logique};
\node[text width=5cm,text centered] at(6,3){\bf  Calculs intermédiaires};
\node[text centered] at(1,3){\bf \large 1};
\node[text centered] at(2,3){\bf \large 1};
\node[text centered] at(3,3){\bf \large 0};
\node[text centered] at(9,3){\bf \large 1};
\node[text centered] at(10,3){\bf \large 0};
\node[text width=5cm,text centered] at(6,5.5){\bf\large Ruban programme};
\node[text centered] at(2,5.5){\bf \large 1};
\node[text centered] at(3,5.5){\bf \large 0};
\node[text centered] at(9,5.5){\bf \large 1};
\node[text centered] at(10,5.5){\bf \large 1};
\node[text width=5cm,text centered] at(6,0.5){\bf\large Ruban résultats};
\node[text centered] at(2,0.5){\bf \large 1};
\node[text centered] at(3,0.5){\bf \large 0};
\node[text centered] at(9,0.5){\bf \large 0};
\node[text centered] at(10,0.5){\bf \large 1};
\end{tikzpicture}

\caption{Calculateur universel\label{fig:calcuniv}}
\end{center}

\vspace*{-4mm}
\hrulefill
\vspace*{-4mm}
\end{figure}
%
%
%
Nous appellerons {\em \bf Calculateur Universel} noté CU, une telle version de MTU à 3 rubans. De même que pour les machines 
de {\em Turing}, nous pouvons classer ces calculateurs $C_0, 
C_1,C_2 \ldots , C_n, \ldots $

Si on considère une suite infinie $p $ de $0$ et de $1$ sur le premier ruban, et $s$ 
une suite finie de $0$ et de $1$ sur le troisième ruban (celui des résultats), on note $C_i(p,s)$ la suite 
finie de $0$ et de $1$ qui se trouve sur le ruban de résultats 
quand le calculateur $C_i$ s'arrête pour le programme $p$ et 
la donnée $s$ (quand ça se produit). Dans le cas où $C_i $ ne 
s'arrête pas pour $p$ et $s$ on dit que $C_i(p,s)$ n'est pas 
défini. Lorsque $C_i(p,s)$ est défini, $C_i$ n'a lu qu'un 
nombre fini de symboles de $p$ (évident, sinon il ne s'arrêterait pas). 
Cette suite finie est appelée {\em programme réduit} et notée 
$pr_i(p,s)$ ce qui se lit, \og{} programme réduit du calculateur 
$i$ pour le programme $p$ et la donnée (l'instance) $s $ \fg.

On est fondé à parler alors de programme réduit. En effet, 
considérons la suite binaire infinie $pr|t$ où $pr$ est le 
programme réduit et $t$ une suite binaire infinie arbitraire, 
et $pr|t$ la concaténation de $pr$ avec $t$. Alors 
$C_i(pr|t,s)$ ne dépend pas de $t$. Toutefois on ne peut pour 
autant con\-sidérer $pr$ comme étant le programme minimal au 
sens du nombre de symboles.
\section{La démarche \label{demarche}}
Le premier problème qui se pose est celui de la gestion du \textbf{temps}.
\subsection{Le temps\label{temps}} 
Le temps, {\em deus ex machina} est, pour ce qui nous concerne, irréversible. 

L'écoulement temporel est marqué dans un CU par le fait que les rubans de programme et de résultat sont finis d'un côté et que les têtes de lecture et d'écriture respectivement sur les rubans programme et résultat ne peuvent ni stagner ni revenir en arrière. Il s'agit là d'une différence fondamentale avec la MTU pour le point de vue qui nous intéresse ici, et qui caractérise le CU par rapport à la MTU.
%
%
\section{La machine-{\Large{$\alpha$}}}

Le calculateur universel ne peut représenter les phénomènes dynamiques, c'est-à-dire ceux dont le "programme" varie en fonction du temps et surtout du passé. Nous présentons ci-après la structure de la machine-{\large{$\alpha$}}, modèle générique pour les algorithmes, qu'ils soient naturels ou classqiues. Nous donnons  la structure de ce modèle.
La figure \ref{univers} représente ce que nous appellerons par la suite machine-{\large{$\alpha$}}.
\subsection{Structure de la machine-{\large{$\alpha$}}}
Sans préjuger du \og{}programme système\fg{}, ou {\em système d'exploitation} de la machine-{\large{$\alpha$}} sur lequel nous reviendrons, elle est constituée ici des trois bandes du calculateur Universel classique. Mais, différence importante, elle est munie de quatre têtes d'accès aux bandes programme, travail et résultat au lieu des trois du CU:
\begin{itemize}
\item[\ding{43}] la tête de lecture du programme qu'on trouve également sur le calculateur universel classique (voir figure \ref{fig:calcuniv}) qui ne peut que lire et se déplacer en un seul sens et d'une seule case à la fois, elle est représentée par un triangle rouge marqué \textbf{r} sur la figure \ref{univers}, elle se déplace d'une seule case à la fois, et ne peut jamais stagner;
\item[\ding{42}] la tête d'\textit{écriture sur le ruban programme} qui est représentée sur la figure \ref{univers} par le triangle bleu marqué \textbf{b}, nous y reviendrons;
\item[\ding{42}] la tête d' \textit{écriture} sur le ruban résultat, qui est représentée sur la figure \ref{univers} par un triangle vert marqué \textbf{v}, elle se déplace d'une seule case à la fois, dans un seul sens, elle peut également écrire sur les rubans \og travail \fg{} d'autres machines dans un réseau (envoi de message par exemple, écriture dans des registres, signal lumineux, cri, \ldots );
\item[\ding{43}] la tête de \textit{lecture-écriture} sur le ruban de travail  figure \ref{univers} par un triangle noir marqué \textbf{n}. Cette tête peut lire ou écrire indifféremment et se déplacer dans les deux sens, d'une seule case à la fois ou stagner, comme dans le CU classique.
\end{itemize}

\subsection{La généricité du modèle}
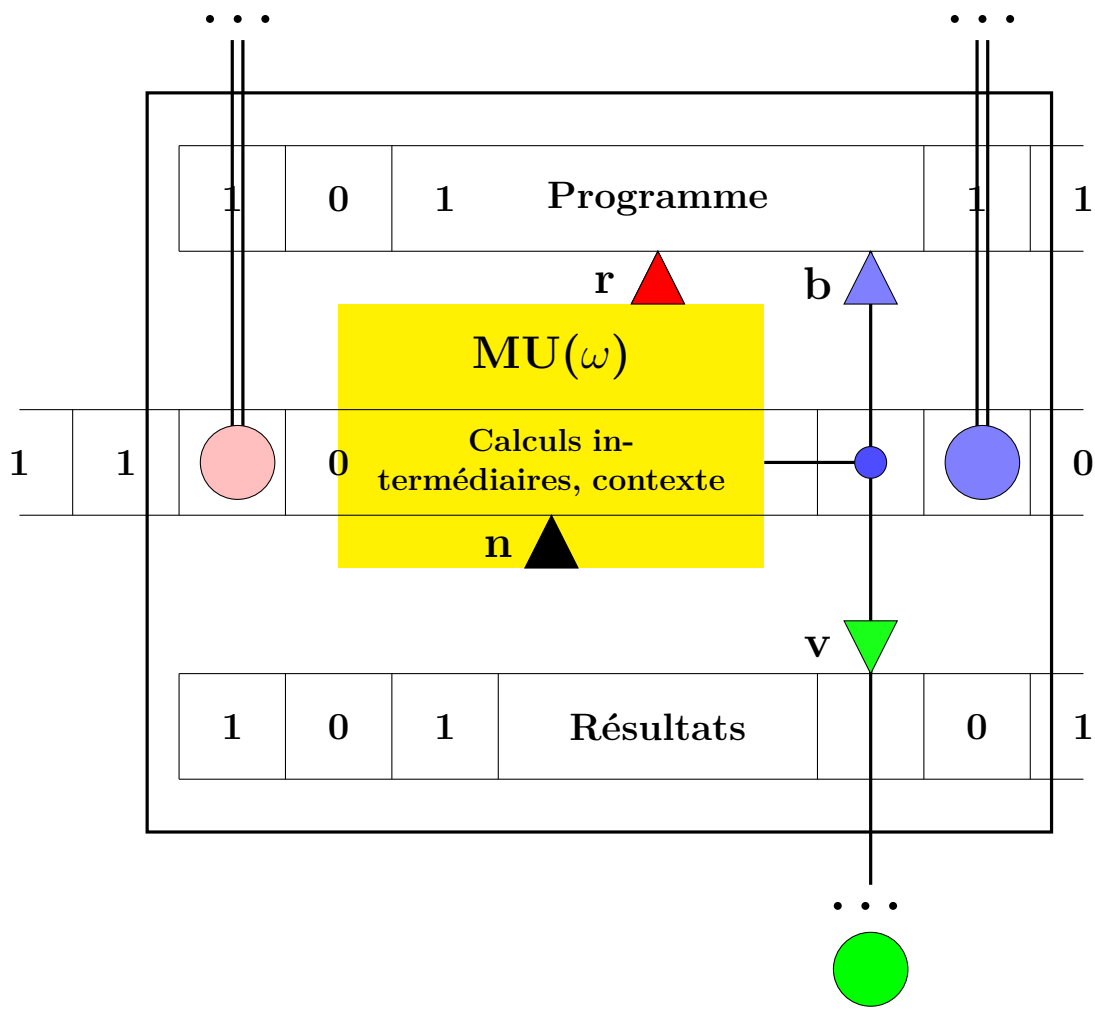
\begin{figure}[h]
\hrulefill

\medskip
\begin{center}
\begin{tikzpicture}[scale=1.4]
\draw[very thick](1.2,-.5) rectangle (9.7,6.5);
\draw(1.5,0)--(10,0);
\draw(1.5,1)--(10,1);
\draw(1.5,1)--(1.5,0);
\draw(2.5,1)--(2.5,0);
\draw(3.5,1)--(3.5,0);
\draw(4.5,1)--(4.5,0);
\draw(7.5,1)--(7.5,0);
\draw(8.5,1)--(8.5,0);
\draw(9.5,1)--(9.5,0);
%
\draw(1.5,5)--(10,5);
\draw(1.5,6)--(10,6);
\draw(1.5,6)--(1.5,5);
\draw(2.5,6)--(2.5,5);
\draw(3.5,6)--(3.5,5);
\draw(8.5,6)--(8.5,5);
\draw(9.5,6)--(9.5,5);
\draw[fill=red!90] (5.75,4.5) -- (6.25,4.5) -- (6,5) -- cycle;
\fill[yellow] (3,2) rectangle +(4,2.5) (5,0); 
%
\draw(0,2.5)--(10,2.5);
\draw(0,3.5)--(10,3.5);
\draw(0.5,2.5)--(0.5,3.5);
\draw(1.5,2.5)--(1.5,3.5);
\draw(2.5,2.5)--(2.5,3.5);
\draw(7.5,2.5)--(7.5,3.5);
\draw(8.5,2.5)--(8.5,3.5);
\draw(9.5,2.5)--(9.5,3.5);
\draw[fill=black] (4.75,2) -- (5.25,2) -- (5,2.5) -- cycle;
\node[text centered] at(4.5,2.2){\textbf{\Large n}};
%
%
\draw(9,3)--(9,7)[very thick];
\draw(9.1,3)--(9.1,7)[very thick];
\draw[black,fill=blue!50] (9.05,3) circle(.35); 
\node[text centered] at(9.1,7.2){\bf \LARGE \ldots};
\draw(2,3)--(2,7)[very thick];
\draw(2.1,3)--(2.1,7)[very thick];
\draw[black,fill=pink!100] (2.05,3) circle(.35); 
\node[text centered] at(2.1,7.2){\bf \LARGE \ldots};
\draw(8,1.5)--(8,-1)[very thick];
\draw[black, fill=green!100] (8,-1.8) circle(.35); 
\node[text centered] at(8,-1.2){\bf \LARGE \ldots};
%
%
\draw(7,3)--(8,3)[very thick];
\draw(8,3)--(8,5)[very thick];
\draw(8,3)--(8,1)[very thick];
\draw[black,fill=blue!70] (8,3) circle(.15); %
\draw[fill=green!90] (7.75,1.5) -- (8.25,1.5) -- (8,1) -- cycle;
\node[text centered] at(7.5,1.25){\textbf{\Large v}};
\draw[fill=blue!50] (7.75,4.5) -- (8.25,4.5) -- (8,5) -- cycle;
\node[text centered] at(7.5,4.7){\textbf{\Large b}};
\draw[fill=red] (5.75,4.5) -- (6.25,4.5) -- (6,5) -- cycle; 
\node[text centered] at(5.5,4.7){\textbf{\Large r}};

%
\node[text width=2.5cm,text centered] at(5,4){\Large \bf MU($\omega$)};
\node[text width=5cm,text centered] at(5,3){\bf Calculs intermédiaires, contexte};
\node[text centered] at(0,3){\bf \large 1};
\node[text centered] at(1,3){\bf \large 1};
\node[text centered] at(3,3){\bf \large 0};
\node[text centered] at(10,3){\bf \large 0};
\node[text width=5cm,text centered] at(6,5.5){\bf\large Programme};%
\node[text centered] at(2,5.5){\bf \large 1};
\node[text centered] at(3,5.5){\bf \large 0};
\node[text centered] at(4,5.5){\bf \large 1};
\node[text centered] at(9,5.5){\bf \large 1};
\node[text centered] at(10,5.5){\bf \large 1};
\node[text width=5cm,text centered] at(6,0.5){\bf\large Résultats};%
\node[text centered] at(2,0.5){\bf \large 1};
\node[text centered] at(3,0.5){\bf \large 0};
\node[text centered] at(4,0.5){\bf \large 1};
\node[text centered] at(9,0.5){\bf \large 0};
\node[text centered] at(10,0.5){\bf \large 1};
%
\end{tikzpicture}
\caption{machine-{\large{$\alpha$}}\label{univers}}
\end{center}

\vspace*{-4mm}
\mbox{}\hrulefill 

\end{figure}

Ce modèle contient le calculateur universel, il suffit d'inactiver la \og{}tête bleue\fg{} (\textit{i.e.} \textbf{b}) d'écriture sur le ruban programme (voir figure \ref{univers}) et de limiter le rôle de la \og{}tête verte\fg{} (\textit{i.e.} \textbf{v}) à l'écriture sur le ruban résultats. En cela ce modèle est déjà plus général que le modèle de CU qui contient lui-même celui de MTU.
 
%
\section{Fonctionnement de la machine-{\Large{$\alpha$}}}
Nous examinerons successivement la variabilité à laquelle sont exposés les codes de la machine-{\large{$\alpha$}}, le {\em système d'exploitation}, et comment se font les communications dans ce modèle générique.
\subsection{La variabilité}
Une fonctionnalité intrinsèque des algorithmes naturels c'est la variabilité, à l'image de la matière vivante\footnote{Nous éviterons ici d'entrer dans le débat sur la définition du vivant.}, et plus précisément si on se réfère au modèle \textit{machine-{\large{$\alpha$}}}, la variabilité du programme afférent. Plus particulièrement, cette variabilité peut être due au contexte ou à l'interaction entre entités modélisées par des machine-{\large{$\alpha$}}.

Dans le modèle  \textit{machine-{\large{$\alpha$}}} cette variabilité est due à l'action des têtes \textbf{b} et \textbf{v} (respectivement bleues et vertes)  sur la figure \ref{univers} et modélisée par leur fonctionnement.
\subsection{Le \og système d'exploitation\fg\label{systeme}}
C'est le système d'exploitation de la \textit{machine-{\large{$\alpha$}}} qui rend compte de cette diversité dans la modélisation. En effet, les interactions avec l'environnement sont ici modélisées par des interventions extérieures provoquant des modifications d'écritures sur le ruban des calculs intermédiaires. Ces modifications sont lues -ou pas- par le \og système d'exploitation\fg{} et traitées en fonction du schème de la \textit{machine-{\large{$\alpha$}}}. C'est là que peuvent intervenir des écritures sur les  trois rubans, suivant à la fois le programme du ruban programme et du schème central.

Le schème central ou {\em système d'exploitation} fonctionne de la manière suivante:
\begin{itemize}
\item Lecture par la tête rouge (\textit{i.e.} \textbf{r} sur la figure) d'une instruction sur le ruban Programme de la figure \ref{univers};
\item écriture ou lecture (en fonction de l'instruction lue précédemment) sur le ruban intermédiaire. En cas de lecture sur le ruban intermédiaire; plusieurs cas peuvent se produire en fonction de ce qui est lu:
	\begin{enumerate}
		\item la tête bleue (\textit{i.e.} \textbf{b}) entre en 		action et écrit sur le ruban programme, elle 			modifie ainsi le programme en cours d'exécution,
		\item la tête verte (\textit{i.e.} \textbf{v}) entre en 		action et écrit sur le ruban résultats,
		\item émission d'un message, c'est-à-dire, 			écriture sur le ruban intermédiaire d'une autre 			machine-{\large{$\alpha$}}.
	\end{enumerate}
\item émission d'un message, c'est-à-dire, écriture sur le ruban intermédiaire d'une autre machine-{\large{$\alpha$}}.
\end{itemize}
Nous avons distingué deux cas d'émission de message en fonction du fait que l'une est prévue dans le programme initial du ruban Programme, alors que l'autre est une réaction à un message reçu, à savoir, une lecture sur le ruban intermédiaire d'une écriture faite par une autre machine-{\large{$\alpha$}}.

La gestion de ces envois et réceptions de messages, de leurs destinataires est faite par le programme inscrit sur le ruban programme.
\subsection{La communication}
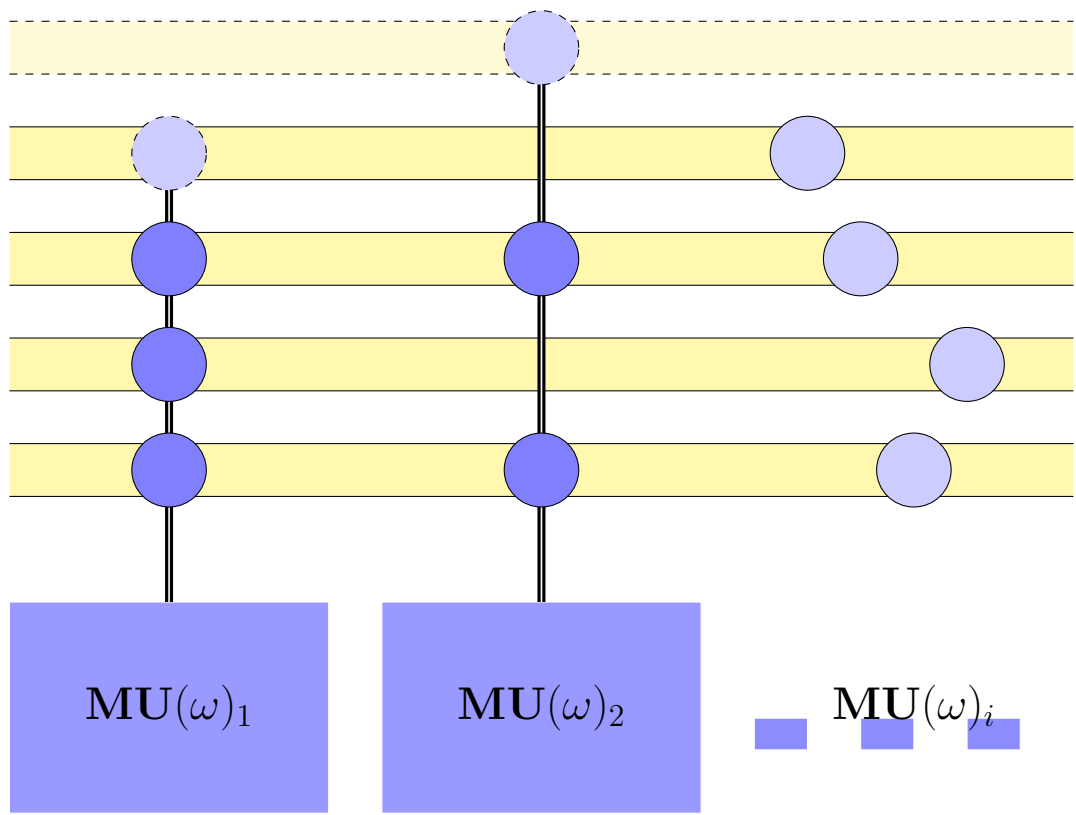
\begin{figure}[h]
\hrulefill

\medskip
\begin{center}
\begin{tikzpicture}[scale=1.4]
\draw[white,fill=yellow!40] (0,3) rectangle (10,3.5);
\draw(0,3)--(10,3);         
\draw(0,3.5)--(10,3.5);   
\draw[white,fill=yellow!40] (0,4) rectangle (10,4.5);
\draw(0,4)--(10,4);       
\draw(0,4.5)--(10,4.5); 
\draw[white,fill=yellow!40] (0,5) rectangle (10,5.5);
\draw(0,5)--(10,5);      
\draw(0,5.5)--(10,5.5);
\draw[white,fill=yellow!40] (0,6) rectangle (10,6.5);
\draw(0,6)--(10,6);
\draw(0,6.5)--(10,6.5);
\draw[white,fill=yellow!20] (0,7) rectangle (10,7.5);
\draw[dashed](0,7)--(10,7);
\draw[dashed](0,7.5)--(10,7.5);

\draw[white,fill=blue!40] (0,0) rectangle (3,2);
\draw[white,fill=blue!40] (3.5,0) rectangle (6.5,2);
\draw[white,fill=blue!45] (7,.6) rectangle (7.5,.9);
\draw[white,fill=blue!45] (8,.6) rectangle (8.5,.9);
\draw[white,fill=blue!45] (9,.6) rectangle (9.5,.9);

%
%
\draw[double,very thick](1.5,2)--(1.5,6);
\draw[double,very thick] (5,2)--(5,7);
%
\draw[dashed,black,fill=blue!20] (1.5,6.25) circle(.35);
\draw[black,fill=blue!50] (1.5,5.25) circle(.35);
\draw[black,fill=blue!50] (1.5,4.25) circle(.35);
\draw[black,fill=blue!50] (1.5,3.25) circle(.35);
\draw[dashed,black,fill=blue!20] (5,7.25) circle(.35);
\draw[black,fill=blue!50] (5,5.25) circle(.35);
\draw[black,fill=blue!50] (5,3.25) circle(.35);
\draw[black,fill=blue!20] (7.5,6.25) circle(.35);
\draw[black,fill=blue!20] (8,5.25) circle(.35);
\draw[black,fill=blue!20] (9,4.25) circle(.35);
\draw[black,fill=blue!20] (8.5,3.25) circle(.35);
\node[text width=2.5cm,text centered] at(1.5,1){\Large \bf MU$(\omega)_{1}$};
\node[text width=2.5cm,text centered] at(5,1){\Large \bf MU$(\omega)_2$};
\node[text width=2.5cm,text centered] at(8.5,1){\Large \bf MU$(\omega)_{i}$};
\end{tikzpicture}
%
\caption{Réseau de \textit{machines-{\large{$\alpha$}}} \label{MTUC}}
\end{center}
\vspace*{-4mm}
\hrulefill
\vspace*{-4mm}
\end{figure}

Le \og{}modèle standard\fg{} de CU ou même de MTU n'est pas prévu pour communiquer. Or les \textit{machine-{\large{$\alpha$}}.} sont connectées en réseau. Ce qui caractérise aussi la vie, c'est l'interaction avec le milieu. Le milieu c'est tout ce qui entoure l'objet d'étude, champs magnétiques, température, oxygène, eau, \ldots ainsi que la communication entre des  \textit{machines-{\large{$\alpha$}}} . La communication peut-être intégrée en considérant le réseau comme une entité. La communication s'opère par l'intermédiaire des rubans (voir figure \ref{MTUC}). Ainsi, par exemple  \textit{machines-{\large{$\alpha$}}}-1 écrit sur le ruban intermédiaire d'une ou plusieurs autres en ce sens comme le schématise la figure \ref{MTUC} qu'on peut considérer que \textit{machine-{\large{$\alpha$}}}-1, noté sur la figure \ref{MTUC} MU$(\omega)_1$ écrivant sur son ruban de travail (ruban intermédiaire sur la figure), une \textit{machine-{\large{$\alpha$}}}-2, MU$(\omega)_{2}$ \textit{machine-{\large{$\alpha$}}}-3, MU$(\omega)_{3}$ \ldots , \textit{machine-{\large{$\alpha$}}}-i, MU$(\omega)_{i}$. Conformément au \S{} \ref{systeme} la communication s'opère par l'intermédiaire du ruban de travail.

On peut ici voir les communications de deux fa\c cons suivant le phénomène modélisé. 

Soit une MU$(\omega)_{i}$ écrit directement sur un ruban de travail d'une autre MU$(\omega)_{j}$;

Soit que toutes les MU($\omega$) ont le même ruban de travail, ce qui implique une synchronisation. 

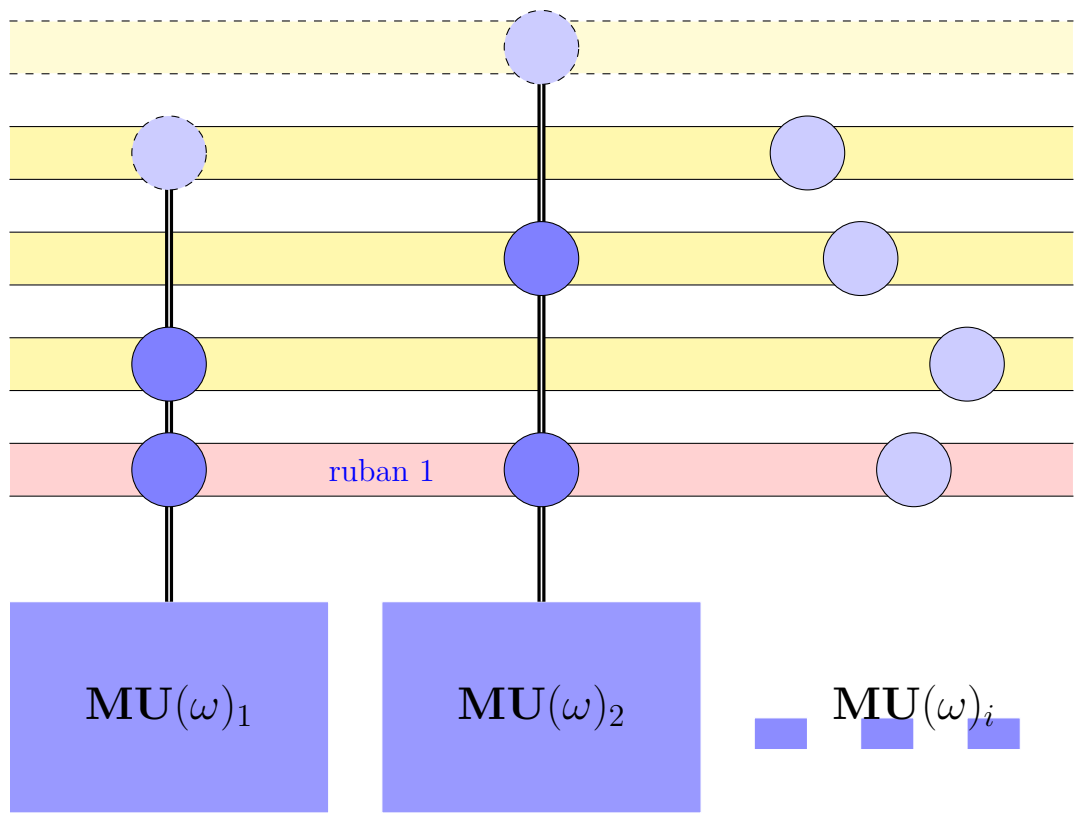
\begin{figure}[h]
\hrulefill

\medskip
\begin{center}
\begin{tikzpicture}[scale=1.4]
\draw[white,fill=pink!70] (0,3) rectangle (10,3.5);
\draw(0,3)--(10,3);         
\draw(0,3.5)--(10,3.5);   
\node[text width=2.5cm,text centered] at(3.5,3.25){\color{blue}{ruban 1}};
\draw[white,fill=yellow!40] (0,4) rectangle (10,4.5);
\draw(0,4)--(10,4);       
\draw(0,4.5)--(10,4.5); 
\draw[white,fill=yellow!40] (0,5) rectangle (10,5.5);
\draw(0,5)--(10,5);      
\draw(0,5.5)--(10,5.5);
\draw[white,fill=yellow!40] (0,6) rectangle (10,6.5);
\draw(0,6)--(10,6);
\draw(0,6.5)--(10,6.5);
\draw[white,fill=yellow!20] (0,7) rectangle (10,7.5);
\draw[dashed](0,7)--(10,7);
\draw[dashed](0,7.5)--(10,7.5);

\draw[white,fill=blue!40] (0,0) rectangle (3,2);
\draw[white,fill=blue!40] (3.5,0) rectangle (6.5,2);
\draw[white,fill=blue!45] (7,.6) rectangle (7.5,.9);
\draw[white,fill=blue!45] (8,.6) rectangle (8.5,.9);
\draw[white,fill=blue!45] (9,.6) rectangle (9.5,.9);
%
%
\draw[double,very thick](1.5,2)--(1.5,6);
\draw[double,very thick] (5,2)--(5,7);
%
\draw[dashed,black,fill=blue!20] (1.5,6.25) circle(.35);
\draw[black,fill=blue!50] (1.5,4.25) circle(.35);
\draw[black,fill=blue!50] (1.5,3.25) circle(.35);
\draw[dashed,black,fill=blue!20] (5,7.25) circle(.35);
\draw[black,fill=blue!50] (5,5.25) circle(.35);
\draw[black,fill=blue!50] (5,3.25) circle(.35);
\draw[black,fill=blue!20] (7.5,6.25) circle(.35);
\draw[black,fill=blue!20] (8,5.25) circle(.35);
\draw[black,fill=blue!20] (9,4.25) circle(.35);
\draw[black,fill=blue!20] (8.5,3.25) circle(.35);
\node[text width=2.5cm,text centered] at(1.5,1){\Large \bf MU$(\omega)_1$};
\node[text width=2.5cm,text centered] at(5,1){\Large \bf MU$(\omega)_2$};
\node[text width=2.5cm,text centered] at(8.5,1){\Large \bf MU$(\omega)_i$};
\end{tikzpicture}
%
\caption{Système de machine-{\large{$\alpha$}} synchronisées\label{MTUCS}}
\end{center}
\hrulefill
\end{figure}

Dans la figure \ref{MTUC} nous avons fait figurer une telle situation; les machines MU$(\omega)_{1}$ et MU$(\omega)_{2}$ sont synchronisées par le ruban de couleur \textcolor{red}{rose}. Dans un tel contexte, le protocole de communication est tel que MU$(\omega)_{1}$ ne peut émettre son message que si MU$(\omega)_{2}$ est en situation de le lire (\textit{i.e.} recevoir).
Les problèmes liés aux systèmes de MU($\omega$) et de communication entre eux, ressortissent à l'algorithmique distribuée de contrôle.
%
%
On pourraît également présenter les choses un peu différemment à partir du même modèle. Dans ce schéma (voir la figure \ref{MTUCS}) le ruban \textit{\color{red}{rose}} modélise une communication entre deux \textit{machines-{\large$\alpha$}}, représentant le fait séquentiel d'une communication qui veut qu'un message ne puisse être lu ou re\c cu avant d'avoir été envoyé.
\section{Des \textit{scenarii}}
\subsection{Le code génétique}
En recodant le code génétique des êtres vivants en binaire, on obtient pour chaque ADN et chaque gène une séquence binaire. Nonobstant le fait que c'est ce qui \og{}construit\fg{}\linebreak  un être vivant, on peut assimiler cette séquence binaire au programme d'un CU. On sait en génétique que nombre de gènes jouent un rôle de \og \textit{codes correcteurs}\fg{} ou ne \og servent à rien\footnote{Il serait plus prudent sans doute de dire qu'on ne sait pas à quoi ils servent ni s'ils servent à quelquechose.}\fg{} on peut par analogie considérer alors la séquence d'ADN comme une séquence binaire au sens de Kolmogorov\footnote{cette façon de faire permet de reconstituer l'arbre philogénétique en considérant les distances entre séquences.} contenant un programme réduit minimal (prm) c'est-à-dire que lors de la construction et même de toute la vie de l'être vivant certains gènes ne sont pas activés, les uns sans doute traduisant un \og{}passé\fg{}, une \og{}généalogie\fg{} de l'organisme et, parmi les autres, des possibles non réalisés, mais potentiellement réalisables sur le long terme.

\subsubsection{Modifications génétiques}
Considérons la variabilité du code génétique. Le code génétique est codé sur le ruban "Programme", lu par la tête {\color{red}rouge}, marquée \textbf{r}; sur le ruban "résultat" est codé l'organisme généré par ce code génétique, inscrit par la tête {\color{green}verte}, marquée \textbf{v}. Le fonctionnement de la  \textit{machines-{\large{$\alpha$}}} MU($\omega_{i}$) influencé aussi par le ruban "contexte" (téte noire) marquée \textbf{n} sur lequel d'autres \textit{machines-{\large{$\alpha$}}} peuvent écrire peut créer des événements en modifiant des symboles sur le ruban programme (tête {\color{blue}bleue} marquée \textbf{b}). La figure \ref{MTUC} représente l'abstraction d'un réseau de  \textit{machines-{\large{$\alpha$}}}  MU($\omega_{i}$), $ i \in \mathbb N^*$ communicant par l'intermédiaire de leurs rubans "contexte\footnote{Nous n'avons pas fait figurer les autres rubans sur la figure pour ne pas l'encombrer.}". On peut d'ailleurs se ramener à un seul ruban \og contexte\fg{} eu égard au fait que ces lectures-écritures sont {\em sérialisables} et que les  \textit{machines-{\large{$\alpha$}}}  sont intrinsèquement séquentiels, cette séquentialité modélisant l'écoulement temporel.

Le code génétique est soumis à variations par le contexte quel qu'il soit, changements électromagnétiques (radiations suite à un accident nucléaire, éruption solaire, modifications du magnétisme terrestre, \ldots), chimiques (Bhopal, agent orange au Vietnam, \ldots), ce qui dans le cas présent sera modélisé par des modifications du codage sur le ruban programme induites par les signaux (messages) re\c cus sur le ruban de travail. 

C'est par un phénomène de modifications du code génétique et de cohérence avec le milieu (la sélection \cite{Dar72}) que s'opère la modification des espèces.

Ajoutons que sur une très longue période de temps (millions d'années par exemple) même des phénomènes de très faible probabilité, éventuellement nulle, finissent par se produire.
\bigskip
\begin{figure}[h]
\hrulefill
\medskip
\begin{center} 
\includegraphics[width=1\linewidth]{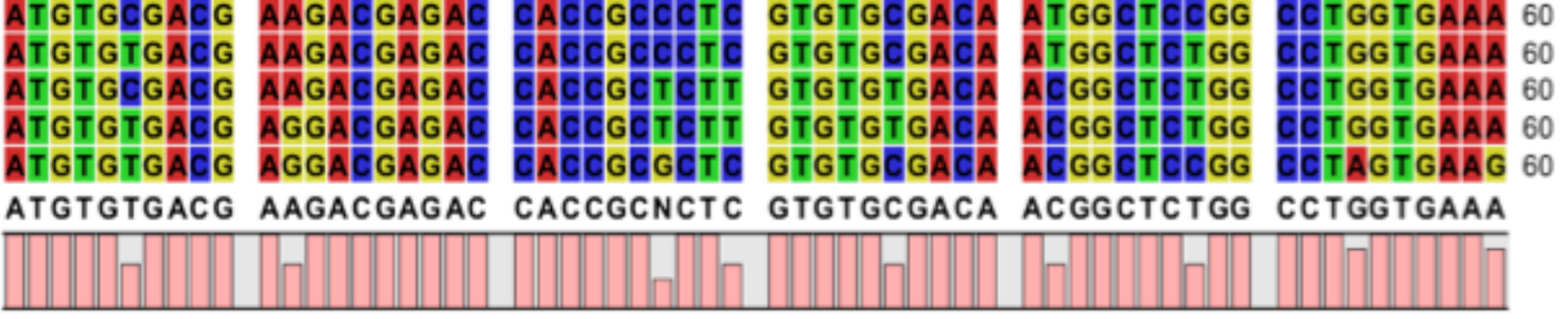} 
\caption{Séquences d'ADN de mamifères\label{kod}} 
\end{center} 

\vspace*{-4mm}
\hrulefill
\end{figure}
\vspace*{-4mm}
\paragraph{Exemple:}
Sur l'exemple de la figure \ref{kod} on a fait figurer les séquences des $60$ premiers nucléides de mamifères (sur 1134), en haut; l'humain, puis dans l'ordre, bovin, souris, rat, poulet. L'histogramme en dessous montre où se situent les modifications, et leur ampleur. Entre le rat et la souris, il y a sur ces $60$ nucléides, deux modifications ($31$ sur les $1134$), le sixième et le onzième, il y en a $120$ sur les $1134$ entre la souris et l'humain, dont $4$ sur les $60$ visibles ici; illustrant ainsi notre propos.
\subsection{L'auto-réplication}
L'auto-réplication qui est une caractéristique du monde biologique est assurée ici, suivant le cas de modélisation voulue, par la recopie du ruban programme sur le ruban résultats dans un cas, ou encore par recopie sur un ruban de travail, qu'il soit celui de la machine-{\large{$\alpha$}} considérée ou d'une autre, soit par synchronisation, soit par message.  C'est au cours de cette réplication que des modifications peuvent apparaître de façon aléatoire comme nous le montrons ci-après au \S  \, \ref{alea}.
\section{Le hasard et la nécessité\label{alea}}
Il est courant dans la littérature anglo-saxonne (par exemple dans \cite{LiVi97}) concernant les modèles théoriques de calcul, d'introduire la notion d'oracle\footnote{Attention, il s'agit la seulement du cas de la théorie de la complexité computationnelle, et il ne faut pas confondre oracle et machine non déterministe. En statistiques par exemple, ce concept peut avoir toute sa place.}, en particulier pour la classe des problèmes $\mathcal{NPC}$\footnote{Il s'agit de la classe d'équivalence de problèmes (plus précisément du problème de {\em satisfiabilité}) pour lesquels on ne connaît pas d'algorithme suceptible de les résoudre en temps polynomial mais dont la vérification d'une solution peut, elle, être faite en temps polyomial.}. Il s'agit alors d'un objet complètement extérieur à la théorie, immanent et artificiel. Nous l'intégrons ici à travers l'introduction d'un événement aléatoire \og {\large{$\omega$}} \fg{} qui gère cet aspect. C'est la modification du ruban de travail due à l'occurence de l'événement {\large$\omega$} d’un événement aléatoire dans le contexte des \textit{machines-{\large{$\alpha$}}} en lieu et place de l’existant.
\subsection{Hasard, contexte et influence} 
Dans la nature, nombres d'événements ressortissent à des phénomènes probabilistes.

En particulier des événements comme ceux qui ont provoqué la disparition d'espèces entières ou des modifications génétiques dans celles existantes, chute de météorites, inversion des pôles magnétiques ressortissent au contexte.Nous introduisons un schéma probabiliste pour modéliser ce type de phénomène. Considérons donc un espace probabilisé $(\Omega, \mathcal A, p)$ décrivant l’univers des possibles $\Omega$, la $\sigma$-algèbre $\mathcal A$ des évènements pouvant se réaliser et $p$ la loi de probabilité définie sur ces évènements.
\begin{itemize}

\item $\Omega$ est un espace abstrait (voir la définition dans \cite{Fre28} , espace des évènements élémentaires en correspondance avec les situations où la machine-{\large{$\alpha$}} va s'exécuter), c'est l'ensemble de tous les événements qui peuvent se produire dans l'environnement;

\item $\mathcal{A}$ est la $\sigma$-algèbre des sous-ensembles mesurables de $\Omega$;

\item et $p($\textbullet$)$ est une mesure de probabilité définie sur $\mathcal A$, $p(E)$ donnant la probabilité d'occurence de $E \in \mathcal{A} $ et satisfaisant $p(\Omega)=1$.
\end{itemize}

La machine-{\large{$\alpha$}}, MU$(\omega)$ modélise donc la machine (son programme, ses données…) dans le contexte $\omega$ où elle est considérée (par exemple, si son code a été modifié par un "orage électromagnétique" qui a changé ses bits de code, son système opératoire (\textit{i.e} O.S.), \ldots .

\begin{definition}[Distance de Leveshtein]\label{dist}
C'est le coût minimal de transformation d'une chaîne de caractères en une autre par utilisation de suppressions, insertion, substitution de caractères à l'une des chaînes. L'algorithme associe une métrique de  coût $1$ à chacune de ces transformations élémentaires portant sur un caractère.
\end{definition}

\subsection{Caractérisation de {\Large $\omega$}}
Lorsqu’un événement $\omega$ se produit, avec une probabilité $p(\omega)$, une modification est engendrée sur le ruban de travail de la \textit{machine-{\large{$\alpha$}}}. Ce ruban est modélisé par un ensemble, noté $E$ de mots de longueur variable construits sur l’alphabet $\{0,1\}$. Si nous notons $m_n, n \in \mathbb N^*$ un mot de longueur $n$, $ E$ s’écrit : $E = \{m_n, n  < + \infty \}$. Cet ensemble $E$ peut être muni d’une métrique telle que la métrique de Levenshtein.

%

Si on note $d_L$ cette distance, $(E, d_L)$ est un espace métrique. On peut alors construire une topologie $T$ sur $E$ qui devient un espace topologique $(E, T)$. L’étape suivante consiste à déterminer la $\sigma-$algèbre $B$ construite à partir de la topologie $T$ définie sur $E$. On obtient ainsi l’espace probabilisable $(E, B)$.
Cela nous amène à considérer l’application $M_n(.)$ définie sur $(\Omega, \mathcal A, p)$ à valeurs dans $(E,B)$ qui à tout $\omega$ de $\Omega$ associe le mot $m_n$ de longueur $n$ généré par $\omega$. Nous postulons maintenant que $M_n(.)$ est une fonction mesurable de $(\Omega, \mathcal A, p)$ dans $(E, B)$, donc une variable aléatoire. 
\begin{rem} 
Si nous supposons que $M_n(.)$ associe à $\omega$, non plus un mot $m_n$ mais un sous-ensemble $B$ de mots de $E$ (de longueurs éventuellement différentes), nous avons la possibilité de modéliser, dans le contexte de la \textit{machine-{\large{$\alpha$}}}, le cas des machines de Turing non déterministes. Il faudra alors considérer des ensembles aléatoires en lieu et place des variables aléatoires.
\end{rem}

Nous avons évoqué précédemment  la prise en compte du temps, notamment dans le cadre de la communication. Sa prise en compte dans l’influence du contexte sur le comportement d’une \textit{machine-{\large{$\alpha$}}} se fait très simplement en considérant, non plus la variable aléatoire $M_n(.)$, mais le processus stochastique $\{Mn,t(.)\}, t≥0$ dans lequel, pour tout $t ≥ 0, M_n,t(.)$ est une variable aléatoire définie sur $(\Omega, \mathcal A, p)$ à valeurs dans $(E, B)$ définie comme $Mn(.)$.
\section{Conclusion}
Nous avons fourni avec  la machine-{\large{$\alpha$}} un modèle générique dont nous avons montré la filiation avec les modèles théoriques de Machine de Turing Universelle, puis de Calculateur Universel et avons caractérisé ce modèle ainsi que la fa\c con dont elle est susceptible de fonctionner sur l'exemple du code génétique.

Cette généralisation de la démarche algorithmique permet d'aborder des problèmes transversaux, ressortissant à des domaines et disciplines scientifiques très divers. De plus, elle permet d'aborder des problèmes foncièrement nouveaux et de mettre en relation, voire d'unifier des problématiques qui sont \textit{a priori} éloignées comme on a pu le voir, qu'il s'agisse de la sélection naturelle, de vols de migrateurs ou de réseaux d'ordinateurs. Il s'agit là d'un nouveau paradigme scientifique.  \\

\hrulefill

\vspace*{-2.5mm}
{\centering  \rule[1pt]{3cm}{.75pt} \par}
\bibliographystyle{apalike}
\bibliography{Biblio_bio.bib}

\hrulefill \\

\textbf{Remerciements} {\small Que soient ici remerciés nos étudiants, les anciens comme les nouveaux doctorants, tous ceux et celles qui, par leurs remarques, leurs thèses en particulier, soutenues ou en cours, nous ont permis de défricher ces idées qui viennent maintenant à maturité mais nous \og travaillent\fg{} depuis déjà quelques années. En particulier il nous faut nommer ici, en ordre alphabétique:
\begin{itemize}
\item Ahat Murat;
\item Ben Amor Soufian;
\item Fouchal Saïd.
\end{itemize}
\end{document}